\documentclass[pre,twocolumn,showpacs,preprintnumbers,amsmath,amssymb]{revtex4-1}

\usepackage{graphicx}
\usepackage{dcolumn}
\usepackage{bm}

\begin{document}

\newcommand{\bea}{\begin{eqnarray}}
\newcommand{\eea}{  \end{eqnarray}}
\newcommand{\bit}{\begin{itemize}}
\newcommand{\eit}{  \end{itemize}}

\newcommand{\be}{\begin{equation}}
\newcommand{\ee}{\end{equation}}
\newcommand{\ra}{\rangle}
\newcommand{\la}{\langle}
\newcommand{\U}{\widetilde{U}}


\def\bra#1{{\langle#1|}}
\def\ket#1{{|#1\rangle}}
\def\bracket#1#2{{\langle#1|#2\rangle}}
\def\inner#1#2{{\langle#1|#2\rangle}}
\def\expect#1{{\langle#1\rangle}}
\def\e{{\rm e}}
\def\proj{{\hat{\cal P}}}
\def\tr{{\rm Tr}}
\def\H{{\hat H}}
\def\Hdag{{\hat H}^\dagger}
\def\Lop{{\cal L}}
\def\Ehat{{\hat E}}
\def\Edag{{\hat E}^\dagger}
\def\Shat{\hat{S}}
\def\Sdag{{\hat S}^\dagger}
\def\Ahat{{\hat A}}
\def\Adag{{\hat A}^\dagger}
\def\U{{\hat U}}
\def\Udag{{\hat U}^\dagger}
\def\Zhat{{\hat Z}}
\def\Phat{{\hat P}}
\def\Op{{\hat O}}
\def\id{{\hat I}}
\def\x{{\hat x}}
\def\P{{\hat P}}
\def\Px{\proj_x}
\def\Pr{\proj_{R}}
\def\Pl{\proj_{L}}


\title{Quantum Parameter Space of Dissipative Directed Transport}

\author{Leonardo Ermann and Gabriel G. Carlo}
\affiliation{Departamento de F\'\i sica, CNEA, Libertador 8250, (C1429BNP) Buenos Aires, Argentina}
\email{ermann@tandar.cnea.gov.ar, carlo@tandar.cnea.gov.ar}

\date{\today}


\begin{abstract}

Quantum manifestations of isoperiodic stable structures (QISSs) have a crucial role in the current behavior 
of quantum dissipative ratchets. In this context, the simple shape of the ISSs 
has been conjectured to be an almost exclusive feature of the classical system. 
This has drastic consequences for many properties of the directed currents, the most important 
one being that it imposes a significant reduction in their maximum values, thus affecting the attainable 
efficiency at the quantum level. 
In this work we prove this conjecture by means of comprehensive numerical 
explorations and statistical analysis of the quantum states. We are able to describe the 
{\em quantum parameter space} of a paradigmatic system for different values of $\hbar_{\rm eff}$ in great detail. 
Moreover, thanks to this we provide evidence on a mechanism that we call {\em parametric 
tunneling} by which the sharp classical borders of the regions in parameter space become 
blurred in the quantum counterpart. We expect this to be a common property of generic 
dissipative quantum systems.
\end{abstract}

\maketitle

The idea of directed transport \cite{Feynman} proved to be very fruitful, 
and has attracted a huge interest in recent years \cite{Reports}
It can be very briefly defined as transport phenomena in spatial and time periodic systems which 
are not subject to thermal equilibrium. The current appears since  all 
spatiotemporal symmetries leading to momentum inversion are broken \cite{origin}. 
Examples of ratchet models (as they are also usually referred to) 
have found application in many areas of research. Here we will mention just a few, such as biology \cite{biology}, 
nanotechnology \cite{nanodevices}, granular crystals \cite{Granular}, 
and some chemical reactions as isomerization \cite{chemistry}. 
This gives an idea of how different the fields of interest could be.

At the classical level, deterministic ratchets with dissipation 
are generally associated with an asymmetric chaotic attractor \cite{Mateos}.
Quantum ratchets show very rich behavior \cite{QR}. 
In this respect we should mention that cold atoms in optical lattices have been deeply investigated 
from both, the theoretical and experimental points of view \cite{CAexp,AOKR}. This extends also to Bose-Einstein 
condensates, which have been transported by means of quantum ratchet accelerators \cite{BECratchets}, where 
the current has no classical counterpart \cite{purelyQR} and 
the energy grows ballistically \cite{ballistic,coherentControl}.
Within this framework, a dissipative quantum ratchet interesting for cold atoms experiments 
has been introduced in \cite{qdisratchets}. Very recently, the parameter space of the classical 
counterpart of this system has been the object of a detailed study \cite{Celestino, Celestino2}. There it has been 
found that families of isoperiodic stable structures (ISSs are Lyapunov stable islands, generic in the 
parameter space of dissipative systems), have a very important 
role in the description of the currents. Subsequently, the effects of temperature have been included in the 
investigations leading to the determination of resistant optimal ratchet transport in its presence \cite{Manchein}. 

When looking at the quantum counterparts of these structures it has recently been found that in general the QISSs look  
like the quantum chaotic attractors at their vicinity in parameter space \cite{Carlo}. In other words, 
the simple structure of the classical ISSs has been conjectured to be an almost exclusive property of 
the classical system. Just in comparatively few cases the quantum structures are similar to these classically 
simple objects (periodic points in the case of maps). One of the main results of this paper consists of providing with a 
comprehensive proof to this conjecture. For that purpose, we give a complete description of the {\em quantum parameter space} 
for two different $\hbar_{\rm eff}$ values. On the other hand, we show how the regions that can be associated to ISSs families 
become interwoven at the quantum level, blurring their classically sharp borders, and thus 
giving rise to what we call {\em parametric tunneling}. 

The system under investigation is a paradigmatic dissipative ratchet system given by the map \cite{qdisratchets,Manchein, Carlo}
\begin{equation}
\left\{
\begin{array}{l}
\overline{n}=\gamma n + 
k[\sin(x)+a\sin(2x+\phi)],
\\
\overline{x}=x+ \tau \overline{n},
\end{array}
\right.
\label{dissmap} 
\end{equation}
where we have denoted by $n$ the momentum variable conjugated to $x$, $\tau$ being the period of the map 
and $\gamma$ the dissipation parameter. 
These equations describe a particle moving in one dimension 
[$x\in(-\infty,+\infty)$] subjected to the periodic kicked asymmetric potential
\begin{equation}
V(x,t)=k\left[\cos(x)+\frac{a}{2}\cos(2x+\phi)\right]
\sum_{m=-\infty}^{+\infty}\delta(t-m \tau),
\end{equation}
again $\tau$ is the kicking period, having a dissipation parametrized by $0\le \gamma \le 1$.
$\gamma=0$ corresponds to the particle in the overdamped regime and $\gamma=1$ to the 
conservative evolution. The directed transport 
appears due to broken spatial ($a \neq 0$ and $\phi \neq m \pi$) and temporal ($\gamma \neq 1$) 
symmetries, we take $a=0.5$ and $\phi=\pi/2$ in this work. The classical dynamics depends only on the 
parameter $K=k \tau$, which can be directly noticed when introducing the rescaled momentum $p=\tau n$.

The quantum version can be obtained following a standard procedure: 
$x\to \hat{x}$, $n\to \hat{n}=-i (d/dx)$ ($\hbar=1$).
Since $[\hat{x},\hat{p}]=i \tau$, the effective Planck constant 
is $\hbar_{\rm eff}=\tau$. The classical limit corresponds to 
$\hbar_{\rm eff}\to 0$, while $K=\hbar_{\rm eff} k$ remains constant.  
Dissipation can be introduced thanks to the 
master equation \cite{Lindblad} for the density operator $\hat{\rho}$ of the 
system 
\begin{equation}
\dot{\hat{\rho}} = -i 
[\hat{H}_s,\hat{\rho}] - \frac{1}{2} \sum_{\mu=1}^2 
\{\hat{L}_{\mu}^{\dag} \hat{L}_{\mu},\hat{\rho}\}+
\sum_{\mu=1}^2 \hat{L}_{\mu} \hat{\rho} \hat{L}_{\mu}^{\dag}. 
\label{lindblad}
\end{equation}
Here $\hat{H}_s=\hat{n}^2/2+V(\hat{x},t)$ is the system
Hamiltonian, \{\,,\,\} is the anticommutator, and $\hat{L}_{\mu}$ are the Lindblad operators 
given by 
\begin{equation}
\begin{array}{l}
\hat{L}_1 = g \sum_n \sqrt{n+1} \; |n \rangle \, \langle n+1|,\\
\hat{L}_2 = g \sum_n \sqrt{n+1} \; |-n \rangle \, \langle -n-1|,
\end{array}
\end{equation} 
with $n=0,1,...$ and $g=\sqrt{-\ln \gamma}$ (according to the Ehrenfest theorem). 
We have evolved $10^6$ classical random initial conditions 
having $p \in [-\pi,\pi]$ and $x \in [0,2\pi]$ ($<\!p_0\!>=0$) in all cases, and also their quantum 
density operator counterpart in a Hilbert space of dimension $N$.

\begin{figure}
  \includegraphics[width=0.4\textwidth]{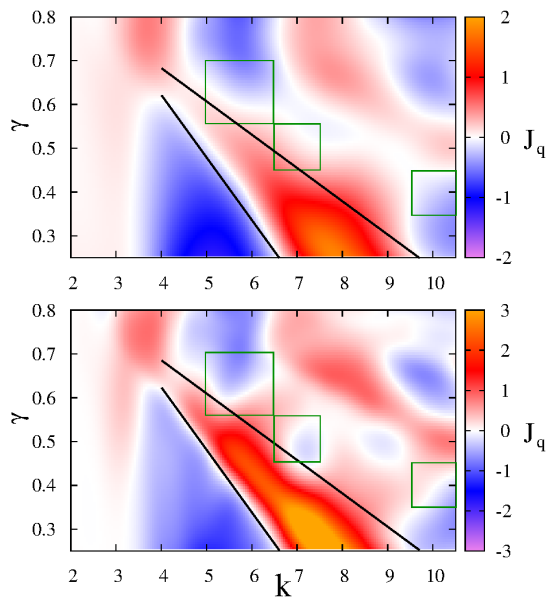} 
 \caption{(color online) Quantum current $J_{\rm q}$ as a function of 
 parameters $k$ and $\gamma$ in a grid of $170\times100$ points.
 The upper panel corresponds to $\hbar_{\rm eff}=0.411$, while the lower one to $\hbar_{\rm eff}=0.137$.
 Black lines correspond to the classical sharp borders of the $B_1$ structure (as in Fig.~1 of \cite{Celestino}).
 (Green) gray squares highlight the regions shown in Fig.~\ref{fig2}.
}
 \label{fig1}
\end{figure}

In \cite{Celestino,Celestino2,Manchein} several classical parameter space portraits have been shown. 
Thanks to them three main kinds of ISSs were identified and called 
$B_M$, $C_M$ and $D_M$, where $M$ stands for an 
integer or rational number and corresponds to the mean momentum of these 
structures in units of $2\pi$. With the exception of $\gamma 
\rightarrow 1$ (i.e., near the conservative limit), ISSs organize the 
parameter space structure and then are essential to understand the current 
behavior. In previous work \cite{Carlo}, sampling a set of relevant points in the quantum parameter space has been 
the only possibility, due to computational restrictions. In this paper we report a major breakthrough in this 
direction, since we were able to completely extend these classical results and provide with quantum parameter space 
portraits in the areas of interest. This can be seen in Fig \ref{fig1}, where we show the quantum current $J_{\rm q}$ 
(we take $J_{\rm q},J_{\rm c}=<\!p\!>$, where $<\!p\!>$ stands for either the quantum or classical average momentum, respectively) 
as a function of  parameters $k$ and $\gamma$. The upper panel corresponds to 
$\hbar_{\rm eff}=0.411$, while the lower one to $\hbar_{\rm eff}=0.137$, both having a 
resolution of $170\times100$ points. They have been obtained by using a cluster having more than $100$ processors. 
It is noticeable how the large $B_1$ structure is almost the only clearly recognizable 
feature that resembles the classical ISSs found in parameter space. There is also a poorly defined region of positive 
current that can be attributed to one of the higher order $B$ families. 
However, the areas associated to chaotic attractors are recognizable, specially the one at $k\sim[3.0,4.0]$ and 
$\gamma\sim[0.6,0.8]$. 

\begin{figure}
  \includegraphics[width=0.4\textwidth]{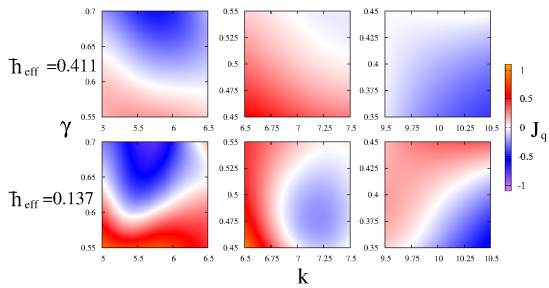} 
 \caption{(color online) Quantum current $J_{\rm q}$ 
 for the three different regions highlighted by means of (green) gray squares in Fig~\ref{fig1}.
 The upper panels correspond to $\hbar_{\rm eff}=0.411$, while the lower ones to $\hbar_{\rm eff}=0.137$, all 
 grids are of $100\times100$ points.}
 \label{fig2}
\end{figure}

In order to investigate what happens with the other ISSs that seem not to have a quantum 
counterpart and also the behavior of the $B_1$ QISS, we explore the three highlighted regions ((green) gray squares) of 
Fig. \ref{fig1} in more detail.
In Fig. \ref{fig2} we show zooms ($100\times100$ points) taken inside these areas, the upper row 
corresponding to $\hbar_{\rm eff}=0.411$ and the lower one to $\hbar_{\rm eff}=0.137$. 
In the left column we can appreciate how the largest of the $C$ structures ($C_{-1}$) influences the $B_1$ region. 
This causes a lowering of the current inside the positive ISS region that is remarkably more pronounced in the $\hbar_{\rm eff}=0.137$ 
case than for $\hbar_{\rm eff}=0.411$. This indicates a kind of tunneling of one structure into the other and since 
this takes place in the parameter space, we propose it as the definition of parametric tunneling. 
Also, there is a positive current chaotic region different from the $B_1$ structure but contiguous to it, that 
forms a continuum with $B_1$ at the quantum level. 
The middle column shows the next zoom area that is connected with 
the previous one by its upper-left corner as can be seen in Fig \ref{fig1} panels. It shares the mentioned  
positive current region and also has a negative current chaotic zone. This latter also 
has an influence on the $B_1$ structure, phenomenon that is clearly more marked for the lower $\hbar_{\rm eff}=0.137$ 
case and that can be appreciated with the help of the Fig. \ref{fig2} middle lower panel. 
Finally, the right column shows an isolated area corresponding to one part of the 
$D_{-1}$ structure which seems to faintly manifest itself through negative currents but whose shape 
makes it difficult to precisely relate it to its classical counterpart. In fact it seems to be 
merged with the negative current chaotic region embedding it.

\begin{figure}
  \includegraphics[width=0.4\textwidth]{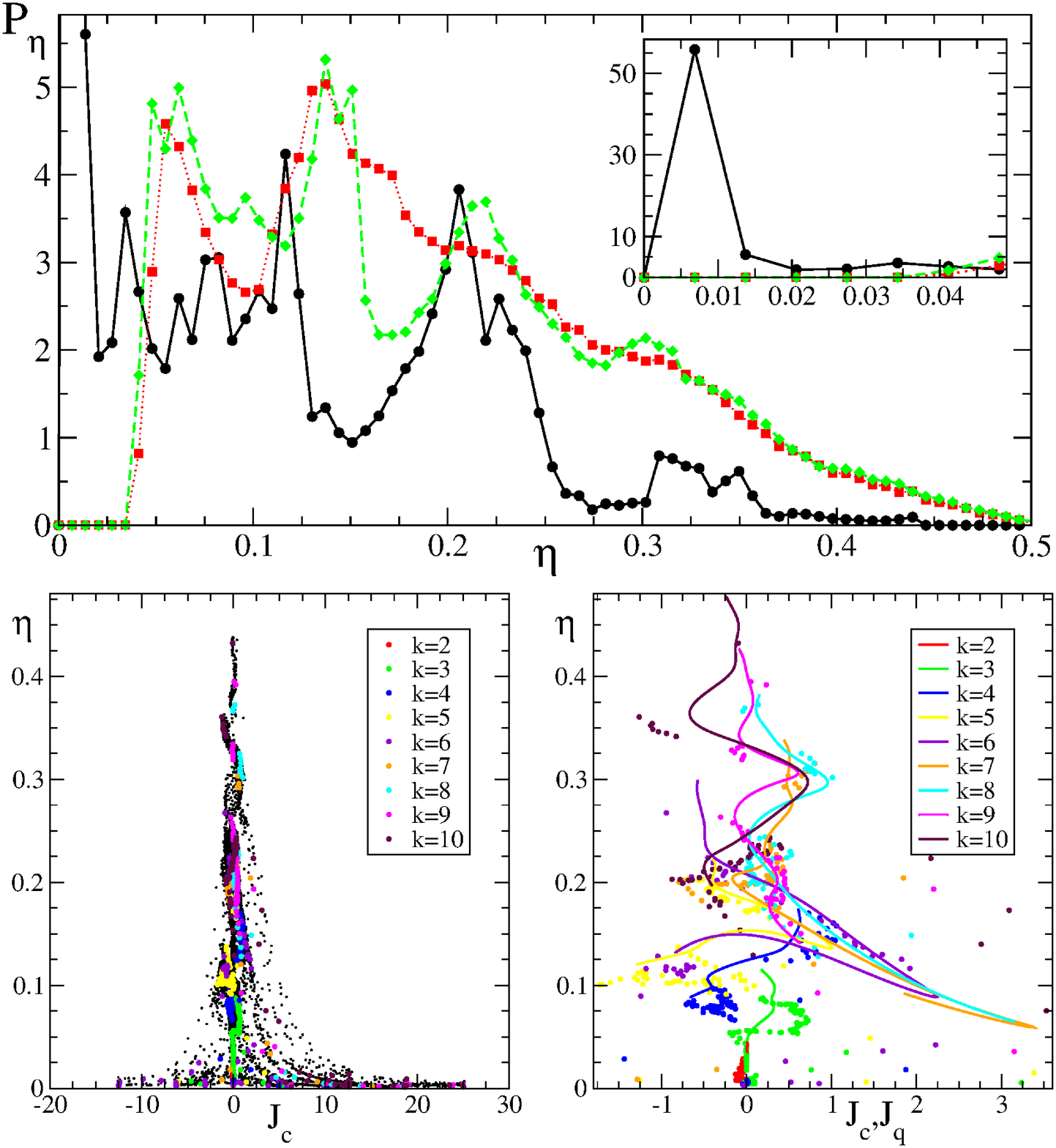} 
 \caption{(color online)
Top panel shows the participation 
ratio histograms $P_{\eta}$ as a function of $\eta$. 
Statistics corresponds to all the cases shown in Fig \ref{fig1}. 
Classical $P_{\eta}$ is represented with black circles (solid line), and 
quantum $P_{\eta}$ for $\hbar_{\rm eff}=0.411$ and $\hbar_{\rm eff}=0.137$
with (red) dark gray squares (dotted line) 
and (green) light gray diamonds (dashed line), respectively.
Inset shows the same histograms in the $\eta \in [0:0.05]$ range.
Bottom left panel: 
black circles represent $\eta$ as a function of $J_{\rm c}$, with 
integer values of $k$ in different (colors) grays.  
Bottom right panel:
comparison of classical (circles) and quantum (lines) 
$\eta$ as a function of $J_{\rm q},J_{\rm c}$ for $\hbar_{\rm eff}=0.137$ with 
integer values of $k=2,\ldots,10$ and 100 values of 
$\gamma\in[0.2,0.8]$. 
  }
 \label{fig3}
\end{figure}

In order to systematically prove that simple (pointlike) structures are exceptional at the quantum level, 
we have analyzed the shape of the limiting quantum momentum distributions (obtained after 50 time steps) by means of 
the participation ratio $\eta=(\sum_iP(p_i)^2)^{-1}/N$. This measure is a good indicator 
of the fraction of basis elements that effectively 
expands the quantum state. For comparative purposes we have also calculated the corresponding classical $\eta$ by 
taking a discretized $p$ distribution (after $10000$ time steps), having a number of bins given by the Hilbert space dimension of 
the lower $\hbar_{\rm eff}$ case (this being $N=3^6$). It is clear that a finer coarse-graining would slightly
change the classical $\eta$ distributions but this will not affect their main properties. This is because 
the distance among points of the ISSs is almost always greater than the chosen bin size. 
We have calculated the histograms $P_{\eta}$ vs. $\eta$,  
and we have also studied how $\eta$ behaves as a function of the current $J_{\rm q},J_{\rm c}$.
Results are shown in Fig. \ref{fig3}, where the upper panel corresponds to the histograms (normalized to 1) for all 
the cases shown in Fig \ref{fig1}. The classical $P_{\eta}$ values (black circles and solid line) have a peak at extremely 
low $\eta$, which can be seen in the inset. The quantum $P_{\eta}$ for $\hbar_{\rm eff}=0.411$ ((red) dark gray squares and dotted line) 
and $\hbar_{\rm eff}=0.137$ ((green) light gray diamonds and dashed line) have larger values at the tail of the distributions. 
These are the most important properties and they are enough to prove our conjecture. Moreover, with the exception of the peak 
around $\eta=0.2$ the classical distribution falls below $0.01$ very quickly, while the quantum ones acquire finite 
values after $\eta\sim0.03$. On the other hand the quantum distribution for $\hbar_{\rm eff}=0.137$ starts to follow 
the shape of the classical one for higher $\eta$ but it is almost unchanged with respect to $\hbar_{\rm eff}=0.411$ 
for the lowest values. 
In Fig. \ref{fig3} bottom left panel we can see that the maximum $J_{\rm c}$ correspond to the lowest $\eta$, and when 
superimposed to the quantum values (see Fig. \ref{fig3} bottom right panel) they roughly match them for some of the 
lowest $J_{\rm q},J_{\rm c}$. In general the classical $\eta$ are extremely discontinuous, a signature of the classical sharp borders 
of the different regions in parameter space, while the quantum ones behave smoothly, also reflecting the 
appearance of the corresponding quantum parameter space. It is worth mentioning that the maximum of $J_{\rm q}$ is 
attained for one of the lowest $\eta$ values and that can be associated to the ``core'' of the $B_1$ region (around 
$k=7.5$ and $\gamma=0.3$). The few lower quantum $\eta$ values have also much lower $J_{\rm q}$ and belong to the lowest $k$ 
region. 

\begin{figure}
  \includegraphics[width=0.4\textwidth]{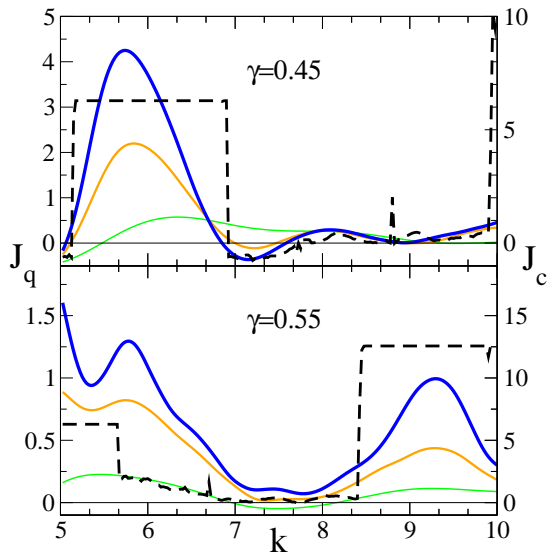} 
 \caption{(color online) Quantum and classical current $J_{\rm q}$ and $J_{\rm c}$ 
 as a function of parameter $k$ for $\gamma=0.45$ (top panel), and $\gamma=0.55$ (bottom panel).
 Quantum currents are represented with solid (green) light gray lines for $\hbar_{\rm eff}=0.411$,
 (orange) gray lines for $\hbar_{\rm eff}=0.137$, and (blue) dark gray lines for $\hbar_{\rm eff}=0.068$. 
 Their corresponding scales are shown on the left side. Classical currents are illustrated with dashed 
 black lines with their corresponding scales on the right. 
}
 \label{fig4}
\end{figure}

Finally, we have selected transversal cuts 
of Fig. \ref{fig1} for two different $\gamma$ values, these being $0.45$, and $0.55$. This shows that the classical 
features are very slowly approached by the quantum distributions and also how different parameter space neighboring 
regions influence each other. The results are shown in Fig. \ref{fig4} upper and 
bottom panels respectively, where it is important to notice that the classical current scale 
(on the right side) is much larger than the quantum one (on the left side). 
It is clearly visible that the sharp borders corresponding to the classical regions cannot 
be reproduced by the quantum counterparts, even though the current significantly grows as $\hbar_{\rm eff}$ drops 
from $0.411$ to $0.068$ for the $B$ QISSs. Moreover we see how negative $J_{\rm q},J_{\rm c}$ values, typical of the chaotic 
region near $B_1$ that is highlighted in the middle panels of Fig. \ref{fig2}, also can be found inside 
a small portion of the $B_1$ area near its border (see Fig. \ref{fig4} upper panel). The influence suffered 
from the $C_{-1}$ structure near $B_1$ can be appreciated as a clear drop in $J_{\rm q}$ values inside the region 
corresponding to this structure (see Fig. \ref{fig4} bottom panel). Again, it is remarkable that this phenomenon 
is more pronounced for the lowest $\hbar_{\rm eff}=0.068$ value.

To summarize, we have performed a comprehensive exploration of 
the quantum parameter space of a paradigmatic system in the directed transport, open systems and 
quantum chaos literature, i.e. the dissipative quantum kicked rotator with a biharmonic kick. 
As a result we were able to develop a complete picture of the {\em quantum parameter space} 
for this kind of systems.
Thanks to statistically exploring the limiting distributions through the participation ratios $\eta$, we have 
systematically proved that the simple structure of the classical ISSs is exceptional at the quantum level. 
This is of huge relevance for the properties of quantum directed currents, 
mainly limiting their maximum values and as such, reducing the efficiency.  
Moreover, we have found a remarkable feature of QISSs, this being the influence that they suffer among 
each other that blurs their classically sharp borders, a phenomenon we have called {\em parametric tunneling}. 
In the future, we will try to identify the details behind this mechanism and also its consequences 
at the current level. Also, we plan to consider a non zero temperature in order to test 
how robust is the obtained quantum parameter space picture.

\vspace{3pc}

Financial support form CONICET is gratefully acknowledged.



\begin{thebibliography}{99}

\bibitem{Feynman}
R. P. Feynman, {\it Lectures on Physics}, {\bf Vol. 1},
(Addison-Wesley,  Reading, MA, 1963).

\bibitem{Reports}
P. Reimann, Phys. Rep. {\bf 361}, 57 (2002); 
S. Kohler, J. Lehmann, and P. H\:anggi, Phys. Rep. {\bf 406}, 379 (2005); 
S. Denisov, S. Flach, and P. H\:anggi, Phys. Rep. {\bf 538}, 77 (2014).

\bibitem{origin}
S. Flach, O. Yevtushenko, and Y. Zolotaryuk, Phys. Rev. Lett. {\bf 84}, 2358 (2000).

\bibitem{biology}
G. Mahmud \emph{et al.}, Nature Phys. {\bf 5}, 606 (2009); 
G. Lambert, D. Liao, and R.H. Austin, Phys. Rev. Lett. {\bf 104}, 168102 (2010);
S. Cocco, J.F. Marko, and R. Monasson, Phys. Rev. Lett. {\bf 112}, 238101 (2014).

\bibitem{nanodevices}
R. D. Astumian, Science {\bf 276}, 917 (1997);
D. Reguera, A. Luque, P.S. Burada, G. Schmid, J.M. Rub\'\i ,and P. H\"anggi, 
Phys. Rev. Lett. {\bf 108}, 020604 (2012).

\bibitem{Granular}
V. Berardi, J. Lydon, P.G. Kevrekidis, C. Daraio, and R. Carretero-Gonz\'alez, 
Phys. Rev. E {\bf 88}, 052202 (2013).

\bibitem{chemistry}
G. G. Carlo, L. Ermann, F. Borondo, and R. M. Benito, Phys. Rev. E {\bf 83}, 
011103 (2011);
L.S. Brizhik, A.A. Eremko, B.M.A.G. Piette, and W.J. Zakrzewski, 
Phys. Rev. E {\bf 89}, 062905 (2014).

\bibitem{Mateos}
J. L. Mateos, Phys. Rev. Lett {\bf 84}, 258 (2000).

\bibitem{QR}
P. Reimann, M. Grifoni, and P. H\"anggi, Phys. Rev. Lett. {\bf 79}, 10 (1997); 
L. Ermann, G. G. Carlo, and M. Saraceno, Phys. Rev. E {\bf 77}, 011126 (2008); 
L. Ermann, G. G. Carlo, and M. Saraceno, Phys. Rev. E {\bf 79}, 056201 (2009).

\bibitem{CAexp}
T. Salger, S. Kling, T. Hecking, C. Geckeler, L. Morales-Molina, and M. Weitz,
  Science {\bf 326}, 1241 (2009).

\bibitem{AOKR}
T. S. Monteiro, P. A. Dando, N. A. C. Hutchings, and M. R. Isherwood,
  Phys. Rev. Lett. {\bf 89}, 194102 (2002);
G. G. Carlo, G. Benenti, G. Casati, S. Wimberger, O. Morsch, R. Mannella, and E. Arimondo,
  Phys. Rev. A {\bf 74}, 033617 (2006).

\bibitem{BECratchets}
M. Sadgrove, M. Horikoshi, T. Sekimura, and K. Nakagawa,
Phys. Rev. Lett. {\bf 99}, 043002 (2007);
I. Dana, V. Ramareddy, I. Talukdar, and G.S. Summy,
Phys. Rev. Lett. {\bf 100}, 024103 (2008);
 D.H. White, S.K. Ruddell, and M.D. Hoogerland
Phys. Rev. A {\bf 88}, 063603 (2013).

\bibitem{purelyQR}
E. Lundh and M. Wallin, Phys. Rev. Lett. {\bf 94}, 110603 (2005);
D. Poletti, G. G. Carlo, and B. Li, Phys. Rev. E {\bf 75}, 011102 (2007).

\bibitem{ballistic}
A. Kenfack, J. Gong, and A.K. Pattanayak, Phys. Rev. Lett. {\bf 100},
044104 (2008); J. Wang and J. Gong, Phys. Rev. E {\bf 78}, 036219 (2008).

\bibitem{coherentControl}
M. Sadgrove, M. Horikoshi, T. Sekimura, and K. Nakagawa,
Eur. Phys. J. D {\bf 45}, 229 (2007).

\bibitem{qdisratchets}
G. G. Carlo, G. Benenti, G. Casati, and D.L. Shepelyansky,
Phys. Rev. Lett. {\bf 94}, 164101 (2005).

\bibitem{Celestino}
A. Celestino, C. Manchein, H.A. Albuquerque, and M.W. Beims, 
Phys. Rev. Lett. {\bf 106}, 234101 (2011).

\bibitem{Celestino2}
A. Celestino, C. Manchein, H.A. Albuquerque, and M.W. Beims, 
Commun. Nonlinear Sci. Numer. Simulat. {\bf 19}, 139 (2014).

\bibitem{Manchein}
C. Manchein, A. Celestino, and M.W. Beims, 
Phys. Rev. Lett. {\bf 110}, 114102 (2013).

\bibitem{Carlo}
G. G. Carlo, 
Phys. Rev. Lett. {\bf 108}, 210605 (2012).

\bibitem{Lindblad}
G. Lindblad, Commun. Math. Phys. {\bf 48}, 119 (1976).

\end{thebibliography}
\end{document}